\documentstyle[aps]{revtex}

\begin{document}
\author{A.E.Meyerovich, S.Stepaniants}
\address{Department of Physics, University of Rhode Island,\\
Kingston, RI 02881}
\title{Transport in Channels and Films with Rough Surfaces. II. Quantized Motion of
Ballistic Particles}
\date{October 18, 1995}
\maketitle

\begin{abstract}
This is the second in the series of papers on transport phenomena along
random rough surfaces. We apply our simple general approach\cite{r1} to
transport in very narrow channels, when the particles wavelength is
comparable to the width of the channels and the motion across the channel is
characterized by discrete quantum states. The discrete nature of the
spectrum leads to a non-analyticity of transport coefficients as a function
of thickness or channel width, especially for degenerate fermions. Surface
inhomogeneity leads to both on-level scattering and interlevel transitions.
As in\cite{r1}, transport coefficients are expressed explicitly via the
correlation function of surface inhomogeneities. The shape of the curves for
the dependence of transport coefficients on the number of particles and/or
film thickness is determined by the correlation radius of surface
inhomogeneities and is not sensitive to their amplitude. For short range
correlations, the curves assume a saw-like shape as a result of the
interlevel transitions. With increasing correlation radius, the interlevel
transitions loose their importance, and the saw teeth gradually decrease,
reducing, in the end, to simple kinks on practically monotonic curves.
Applications include Gaussian and short-range $\delta $-type correlations.
Careful analysis of the transition from quantum to semi-classical and
classical regimes allowed us to improve the accuracy of our previous
classical calculations.

\end{abstract}

\section{Introduction}

Repeated collisions of (ballistic) particles with rough walls with random
inhomogeneities result in gradual chaotization of motion of particles along
the walls, and lead to formation of the mean free path. Though this is a
very old problem, important for different branches of physics, a simple
consistent description has been absent. Recently we suggested a new simple
quantitative approach to transport phenomena in systems with rough walls\cite
{r1} (referred below as {\rm I}; see also\cite{r2}). Within this approach,
one can easily express transport and localization parameters (such as
mobility, diffusion, mean free path, localization length, {\it etc.}) for
the motion of ballistic particles along the walls directly via the
correlation function of wall inhomogeneities. The only restriction is that
the inhomogeneities should be relatively smooth with the amplitude $\ell $
smaller than their correlation radius $R$ and the thickness of the film
(width of the channel) $L$, $\ell \ll L,R$ .

Our approach is based on the use of an explicit canonical coordinate
transformation (similar to the Migdal transformation in nuclear physics)
which makes the initially rough boundaries flat. The price is a considerable
complication of the bulk transport equation which, however, can be treated
perturbatively. This transformation corresponds to stretching of the film
(channel) width with the stretching parameters changing along the film in
accordance with the exact shape of the boundaries. Since, in contrast to
earlier, more heuristic approaches\cite{r3,r4}, we are using an explicit
expression for the coordinate transformation with the parameters given by
the exact profile of the random boundaries, our reformulation of the
transport problem with random rough walls as the transport problem with flat
walls and randomly distorted bulk, is exact. The latter problem allows a
straightforward perturbative solution which involves averaging over initial
boundary inhomogeneities.

In many cases, all the calculations after the coordinate transformation are
really very simple and straightforward. In {\rm I} we illustrated our method
for a wide range of the problems. In this paper we will concentrate on the
case of very thin films and narrow channels when the motion of (longwave)
particles across the channel is quantized with a noticeable separation
between the states.

One of the effects that we want to describe is the following. In thin films
with discrete levels for motion across the film, the increase in particle
density and/or film thickness cause redistribution of particles between
levels. In Fermi systems at $T\rightarrow 0$ this is a non-analytical
step-like process. In this case, the lower levels are filled, while the
higher levels are unoccupied. The (gradual) change in the total number of
particles results, at certain critical values, in the change of the number
of occupied levels by one. These abrupt changes in the number of occupied
levels should lead to a non-analytical dependence of the density of states
and other thermodynamic functions on the number of particles. Obviously,
this effect should also lead to singularities in the dependence of the
transport coefficients on film thickness or density of particles. This
saw-like effect has already been described in the case of bulk scattering%
\cite{r6}, and has been qualitatively suggested in Ref.\cite{r4} for
scattering by rough walls. We will be able to get quantitative results, and
express explicitly the transport singularities via the correlation function
of surface inhomogeneities.

What is more, we will demonstrate that the picture is even more interesting
than in\cite{r4,r6}. We will see that the particle transport along the film
is a non-trivial function of two parameters, $NL^2$ and $L/R$, where $L$ is
the thickness of the film with the $2D$ density of particles $N$, and $R$ is
the correlation radius of surface inhomogeneities. The saw-like dependence
of the transport coefficients on $NL^2$ is well-pronounced only for
short-range correlations of surface inhomogeneities, $NR^2\ll 1$. With an
increase of the correlation radius $R$, the saw-like structure of the curves
will gradually disappear, and the curves will become more and more smooth.
It turns out that the saw-like shape of the curves is associated exclusively
with the interlevel transitions the rate of which is determined by the
parameter $NR^2$; if one artificially forbids the interlevel transitions
even at small $NR^2$, the curves will loose their saw-like character and
will become similar to those at large $NR^2$.

Of course, the density dependence of transport coefficients becomes more and
more smooth with increasing temperature even though the energy spectrum
remains distinctly discrete.

In the next Section we will present general equations for quantized
transport in thin films with rough boundaries. Then, in Sec.3, we will study
transport singularities for degenerate fermions at $T=0$. In Sec.4 we will
derive the corresponding transport equations for finite temperatures, and
calculate the transport coefficients in Boltzmann temperature range. Our
detailed analysis of transport of particles with discrete spectrum will
demonstrate that the accuracy of previous calculations for continuous
spectrum ({\it i.e.,} in classical and/or semi-classical limit) in {\rm I}
can be improved considerably. The consistent transition from discrete to
continuous expressions will provide us with much more accurate expressions
for the squares of the $\delta $-functions, which appear in the calculations
of the squares of the matrix elements, than the approximations used in
purely classical calculations {\rm I}. The improved classical transport
calculations will be described in the Appendix.

\section{Quantized States Across the Channels}

As in {\rm I}, we will consider a film (channel) of the average thickness $L$
with rough boundaries $x=L/2-\xi _1(y,z)$ and $x=-L/2+\xi _2(y,z).$ The
small boundary inhomogeneities, $\xi _1,\xi _2\ll L$, are random functions
of coordinates ${\bf s=}\left( y,z\right) $ along the boundaries, $%
\left\langle \xi _1\right\rangle =\left\langle \xi _2\right\rangle =0$, with
the correlation function 
\begin{eqnarray}
\zeta _{ik}\left( \left| {\bf s}_1-{\bf s}_2\right| \right) &=&\left\langle
\xi _i({\bf s}_1)\xi _k({\bf s}_2)\right\rangle ,  \label{a1} \\
\zeta _{ik}\left( {\bf q}\right) &=&\int d^2s\ e^{i{\bf q\cdot s}/\hbar
}\zeta _{ik}\left( {\bf s}\right)  \nonumber
\end{eqnarray}
where ${\bf q}$ is the momentum along the wall. In homogeneous systems, the
correlation function depends only on the distance between points $\left| 
{\bf s}_1-{\bf s}_2\right| $ and not on coordinates themselves. In the
absence of the bulk relaxation, the results, as in {\rm I}, depend only on
the function $\xi \left( {\bf s}\right) =\xi _1\left( {\bf s}\right) +\xi
_2\left( {\bf s}\right) $ and the correlation function $\zeta \left| {\bf s}%
_1-{\bf s}_2\right| =\left\langle \xi ({\bf s}_1)\xi ({\bf s}%
_2)\right\rangle =\zeta _{11}+\zeta _{22}+2\zeta _{12.}$

Though we can calculate the transport coefficient for arbitrary correlation
function $\zeta \left( s\right) $, we will supplement general expressions by
the most practical examples of Gaussian correlations of the surface
inhomogeneities of an average height $\ell $, 
\begin{equation}
\zeta \left( s\right) =\ell ^2\exp \left( -s^2/2R^2\right) ,\ \ \zeta \left( 
{\bf q}\right) =2\pi \ell ^2R^2\exp \left( -q^2R^2/2\hbar ^2\right)
\label{a2}
\end{equation}
including the limiting case of the very small correlation radius $R$, {\it %
i.e.} the $\delta $-type correlations, 
\begin{equation}
\zeta \left( s\right) =\ell ^2R^2\delta \left( s\right) /s,\ \ \zeta \left( 
{\bf q}\right) =2\pi \ell ^2R^2  \label{a3}
\end{equation}
Note, that the condition $\ell \ll R$ does not mean that our approach is
applicable to long-range correlations (large size inhomogeneities)
exclusively. The scale for the effective correlation range in Eq.$\left( 
\text{\ref{a2}}\right) $ is defined by the particles wavelength $\lambda $.
For longwave particles $\lambda \gg R$ one deals effectively with the
short-range $\delta $-type correlations $\left( \text{\ref{a3}}\right) $,
while in the opposite case of long-range correlations $\lambda \ll R$ one
should consider the full Gaussian expression $\left( \text{\ref{a2}}\right) $%
.

Our approach is based on the use of canonical coordinate transformation 
\begin{equation}
X=\frac{L[x-\frac 12(\xi _2(y,z)-\xi _1(y,z))]}{L-(\xi _1(y,z)+\xi _2(y,z))},%
\text{ }Y=y,\text{ }Z=z  \label{a4}
\end{equation}
which corresponds to the following change of the form of the bulk
Hamiltonian $\widehat{H}=p^2/2m$: 
\begin{equation}
\widehat{H}=\frac{\widehat{P}^2}{2m}+\widehat{V},\text{ }\widehat{V}=\frac 
\xi {mL}\widehat{P}_x^2+\frac 1{2m}\left( X\widehat{P}_x\frac{\xi _y^{\prime
}}L\widehat{P}_y+X\widehat{P}_x\frac{\xi _z^{\prime }}L\widehat{P}%
_z+H.c.\right)  \label{a5}
\end{equation}
(for details see {\rm I}). The randomness of inhomogeneities, $\left\langle
\xi \right\rangle =0,$ leads to the randomness of the bulk ''perturbation'' $%
\widehat{V}$, $\left\langle \widehat{V}\right\rangle =0$. Thus, the
transformation $\left( \text{\ref{a4}}\right) $ reduces the transport
problem between rough walls to an equivalent transport problem with ideal
specular walls, $\Psi (L/2)=\Psi (-L/2)=0$, but with a distorted bulk
Hamiltonian $\left( \text{\ref{a5}}\right) $. The latter problem can be
treated in the same standard perturbative way as for any random bulk
imperfections or impurities.

In thin films, the motion of particles across the films is quantized, $%
P_{x_j}=\pi j/L$, while the motion along the films can still be considered
classically or semi-classically. Then the wave functions which correspond to
the unperturbed Hamiltonian $\left( \text{\ref{a5}}\right) $ with $\widehat{V%
}=0$ have the form 
\begin{equation}
\Psi _j\left( {\bf r}\right) =\sqrt{2/v_0}\exp \left( i{\bf q\cdot s}\right)
\sin \left( \pi jX/L\right)  \label{a6}
\end{equation}
($v_0$ is the volume). In {\rm I} we discussed the simplest situation when
the distance between levels $\left( \text{\ref{a6}}\right) $ is so large
that the transition between them are practically forbidden. Then the
transport problem reduces to independent calculations of the mean free paths 
${\cal L}_j$ for each level. The latter problem was treated as classical $2D$
diffusion on each level.

The full quantum problem should include the transition between levels $%
\left( \text{\ref{a6}}\right) $ with different $j$ caused by the
perturbation $\left( \text{\ref{a5}}\right) $. Surface inhomogeneities lead
to the change of quantized energy levels $\epsilon _{j{\bf Q}},$ interlevel
transitions $j\rightarrow j^{\prime }$ with diffusion between energy levels,
and diffusion of particles on the same level. Since the perturbation is
small and we are interested only in {\it transport} along the film, we can,
as in {\rm I}, neglect perturbation-induced corrections to the energy levels
and neglect the distortion $\left( \text{\ref{a5}}\right) $ everywhere
except for the collision integral in the perturbative transport equation (%
{\it i.e., }neglect all distortions in dynamic, {\it l.h.s.} of the
transport equation). Then the transport equation becomes a set of equations
in the distribution functions $n_j$ coupled via collision integrals $L_j$: 
\begin{equation}
\partial _tn\left( \epsilon _j,{\bf q}\right) +\frac{{\bf q}}m\cdot \partial
_{{\bf r}}n\left( \epsilon _j,{\bf q}\right) +{\bf F\cdot \partial }_{{\bf q}%
}n\left( \epsilon _j,{\bf q}\right) =L_j\left\{ n_i\right\}  \label{a7}
\end{equation}
where $\epsilon _j=\left[ \left( \pi j\hbar /L\right) ^2+q^2\right] /2m$.
The collision integrals $L_j$ are determined by the transition probabilities 
$W_{jj^{\prime }}\left( {\bf q,q}^{\prime }\right) $between states $\left( j,%
{\bf q}\right) \rightarrow \left( j^{\prime },{\bf q}^{\prime }\right) $, 
\begin{equation}
L_j=\sum_{j^{\prime }}\int W_{jj^{\prime }}\left( {\bf q,q}^{\prime }\right)
\left[ n_j(1-n_{j^{\prime }}^{\prime })-n_{j^{\prime }}^{\prime
}(1-n_j)\right] \frac{d^2q^{\prime }}{\left( 2\pi \hbar \right) ^2}
\label{a8}
\end{equation}
The transition probabilities are given by the squares of the matrix elements
of the perturbation $\left( \text{\ref{a5}}\right) $: 
\begin{equation}
W_{jj^{\prime }}\left( {\bf q,q}^{\prime }\right) =\frac{2\pi }\hbar
\,\left\langle \left| V_{{\bf jq},j^{\prime }{\bf q}^{\prime }}\right|
^2\right\rangle \,\delta \left( \epsilon _j\left( {\bf q}\right) -\epsilon
_{j^{\prime }}^{\prime }\left( {\bf q}^{\prime }\right) \right)  \label{a9}
\end{equation}
Since all the equations include only the squares of the matrix elements of
the ''perturbation'' $\widehat{V}$, the averaging over the random surface
inhomogeneities does not cause any problems leading directly to the
correlation function $\zeta \left( s\right) $.

The matrix elements of the perturbation $\left( \text{\ref{a5}}\right) $ are 
\begin{eqnarray}
V_{{\bf q}j,{\bf q}^{\prime }j^{\prime }} &=&V_{{\bf q}j,{\bf q}^{\prime
}j^{\prime }}^{\left( x\right) }+V_{{\bf q}j,{\bf q}^{\prime }j^{\prime
}}^{\left( y\right) }+V_{{\bf q}j,{\bf q}^{\prime }j^{\prime }}^{\left(
z\right) }  \label{a10} \\
V_{{\bf q}j,{\bf q}^{\prime }j^{\prime }}^{\left( x\right) } &=&\frac{\delta
_{jj^{\prime }}}{2mL}\frac{\pi ^2\left( j^2+j^{\prime 2}\right) }{L^2}\xi
\left( {\bf q-q}^{\prime }\right) ,  \nonumber \\
V_{{\bf q}j,{\bf q}^{\prime }j^{\prime }}^{\left( y\right) } &=&\frac{\left(
-1\right) ^{j+j^{\prime }}}{2mL}\xi \left( {\bf q-q}^{\prime }\right) \left(
q_y^{\prime }-q_y\right) \times  \nonumber \\
&&\left[ q_y^{\prime }j^{\prime }\left( \frac 1{j+j^{\prime }}+\frac 1{%
j-j^{\prime }}\left( 1-\delta _{jj^{\prime }}\right) \right) -q_yj\left( 
\frac 1{j+j^{\prime }}-\frac 1{j-j^{\prime }}\left( 1-\delta _{jj^{\prime
}}\right) \right) \right]  \nonumber
\end{eqnarray}
and the collision integral $\left( \text{\ref{a8}}\right) $ for particles on
each level $j$ becomes, after averaging over inhomogeneities, equal to 
\begin{eqnarray}
L_j &=&\frac 1{2\pi \hbar ^3m^2L^2}\int d^2q^{\prime }\;\zeta \left( {\bf q-q%
}^{\prime }\right) \sum_{j^{\prime }}\left( n_{j^{\prime }}\left( {\bf q}%
^{\prime }\right) -n_j\left( {\bf q}\right) \right) \delta \left( \epsilon
_{j^{\prime }{\bf q}^{\prime }}-\epsilon _{j{\bf q}}\right) \times  \nonumber
\\
&&\left[ \delta _{jj^{\prime }}\left( \frac 14\left( {\bf q-q}^{\prime
}\right) ^2+\left( \frac{\pi \hbar j}L\right) ^2\right) ^2+\frac{\left(
1-\delta _{jj^{\prime }}\right) j^2j^{\prime 2}}{\left( j^2-j^{\prime
2}\right) ^2}\left( q^{\prime 2}-q^2\right) ^2\right] .  \label{a11}
\end{eqnarray}
This collision integral includes both on-level scattering (diagonal terms
with $\delta _{jj^{\prime }}$) and interlevel transitions (off-diagonal
terms with $1-\delta _{jj^{\prime }}$) which are induced by surface
roughness.

In {\rm I} we assumed that the interlevel transitions disappear ({\it i.e.,}
that the off-diagonal terms in $\left( \text{\ref{a11}}\right) $ can be
neglected) with an increase in the distance between levels. We solved the
transport equation without interlevel transition with an additional
assumption that the film is very thin and, therefore, the components of
particle momenta across the film are larger than those along the film, $%
\left( \pi \hbar j/L\right) ^2\gg \left( {\bf q-q}^{\prime }\right) ^2$. The
results below show that these assumptions are self-consistent for large
correlation radii of surface inhomogeneities, $NR^2\gg 1$. For smaller
correlation radii the interlevel transitions and momenta along the film
cannot be neglected even in the case of large interlevel spacing. As we will
see, the approximation {\rm I}, though valid in important limiting cases,
misses the most interesting features of transport through very narrow
channels or thin films.

\section{Singularities in transport of particles with discrete quantum
states: Low temperatures}

Changes in particle density and/or thickness of the film lead to
redistribution of particles between discrete quantum states $j$. This
redistribution between {\em discrete} states may lead to a non-analytic
dependence of transport coefficients on particle density and thickness of
the film. Of course, this non-analyticity is more pronounced for degenerate
Fermi systems at $T\rightarrow 0$ when continuous increase in the number of
particles leads, at certain critical densities, to filling of new levels
with higher and higher values of $j$. Below we will look at this
non-analytic effect in some detail.

At $T=0$, the Fermi momenta of fermions for the motion along the film $%
q_F^{\left( j\right) }$on each level $j$ are given by the overall Fermi
energy (chemical potential) $\epsilon _F$ as 
\begin{equation}
\epsilon _F=\frac 1{2m}\left( \left( \frac{\pi j\hbar }L\right)
^2+q_F^{\left( j\right) 2}\right)  \label{a12}
\end{equation}
while the $2D$ density of spin-1/2 particles on each level is 
\begin{equation}
N_j=\frac{q_F^{\left( j\right) 2}}{2\pi \hbar ^2}  \label{a13}
\end{equation}
(for simplicity we assume that the effective masses of particles on all
levels are the same). The chemical potential $\mu =\epsilon _F$ should, as
usually, be determined self-consistently by calculating the total density of
particles $N$, 
\begin{equation}
N=\sum_jN_j=\frac 1{2\pi \hbar ^2}\sum_j\left( 2m\epsilon _F-\left( \frac{%
\pi j\hbar }L\right) ^2\right) .  \label{a14}
\end{equation}
Eqs. $\left( \text{\ref{a12}}\right) $-$\left( \text{\ref{a14}}\right) $ in
convenient dimensionless notations, 
\begin{equation}
\nu =2m\epsilon _F\left( \frac L{\pi \hbar }\right) ^2,\ z_j=\frac 2\pi
N_jL^2,\ z\equiv \sum_jz_j=\frac 2\pi NL^2,  \label{a15}
\end{equation}
can be rewritten as 
\begin{equation}
z_j=\nu -j^2,\ z=\sum z_j  \label{a16}
\end{equation}
The number of occupied levels $S$ for the given value of $z$ ({\it i.e.},
for the number of particles $NL^2$) is given by the integer part of $\nu
^{1/2}\left( z\right) $, 
\begin{equation}
S\left( z\right) ={\rm Int\,}\left[ \sqrt{\nu }\right]  \label{a17}
\end{equation}
All the levels with the indices $j>S$ are empty, $z_{j>S}=0$. Summation of
Eqs.$\left( \text{\ref{a16}}\right) $ from $1$ to $S$ defines the number of
occupied levels $S$ and the dimensionless chemical potential $\nu $ as
functions of the number of particles $z$: 
\begin{eqnarray}
S &=&{\rm Int\,}\left[ \sqrt{\nu }\right] ={\rm Int\,}\left[ \sqrt{\frac zS+%
\frac{\left( S+1\right) \left( 2S+1\right) }6}\right] ,  \label{a18} \\
\nu \left( z\right) &=&\frac zS+\frac 16\left( S+1\right) \left( 2S+1\right)
\nonumber
\end{eqnarray}
[For computational purposes, it is more convenient to start the calculations
by defining the number of occupied levels $S$, and to determine the interval
of the values of $z$ and $\nu $, which corresponds to this number of levels,
basing on the value of $S$]. The changes in number of occupied levels $S=1,\
2,\ 3,\ 4,\ 5,\ 6,...$ occur at $z=0,\ 3$, $13$, $34,$ $70,$ $125,...$ ({\it %
i.e.,} in the points $z=S^3-S\left( S+1\right) \left( 2S+1\right) /6$).

At $T=0$, we look for the solution of the transport equation $\left( \text{%
\ref{a7}}\right) $ in the form 
\[
n_j\left( {\bf q}\right) =n_j^{\left( 0\right) }\left( q_F^{\left( j\right)
}\right) -\frac{FL^3}{\pi ^4\ell ^2}\delta \left( \epsilon -\epsilon
_F\right) \chi _j\left( q_F^{\left( j\right) }\right) \cos \theta _j,
\]
where $\theta _j$ is the angle between the momentum ${\bf q}_j$ and the
external force ${\bf F}$. Then, after the integration of the collision
integral $\left( \text{\ref{a11}}\right) $ with the Gaussian correlation of
surface inhomogeneities $\left( \text{\ref{a2}}\right) ,$ the transport
equation reduces to the following set of $S$ dimensionless linear equations
in $\chi _j\left( q_F^{\left( j\right) }\right) $: 
\begin{eqnarray}
\frac{z_j^{1/2}L^2}{R^2} &=&-\frac 12\chi _j\times   \label{b1} \\
&&\left( 4j^4\,_1F_1\left( \frac 32,2,-\frac{2\pi ^2z_jR^2}{L^2}\right)
+6z_jj^2\,_1F_1\left( \frac 52,3,-\frac{2\pi ^2z_jR^2}{L^2}\right) +\frac 52%
z_j^2\,_1F_1\left( \frac 72,4,-\frac{2\pi ^2z_jR^2}{L^2}\right) \right)  
\nonumber \\
&&+2\sum_{j^{\prime }}^{S\left( z\right) }\left( 1-\delta _{jj^{\prime
}}\right) j^2j^{\prime 2}\exp \left[ -\pi ^2\left( \sqrt{z_j}-\sqrt{%
z_{j^{\prime }}}\right) ^2R^2/2L^2\right] \times   \nonumber \\
&&\left[ \chi _{j^{\prime }}\left( _1F_1\left( \frac 12,1,-\frac{2\pi ^2%
\sqrt{z_jz_{j^{\prime }}}R^2}{L^2}\right) -\,_1F_1\left( \frac 32,2,-\frac{%
2\pi ^2\sqrt{z_jz_{j^{\prime }}}R^2}{L^2}\right) \right) -\chi
_j\,_1F_1\left( \frac 12,1,-\frac{2\pi ^2\sqrt{z_jz_{j^{\prime }}}R^2}{L^2}%
\right) \right]   \nonumber
\end{eqnarray}
(we will not give here similar cumbersome equations for the correlation
function of a general form $\zeta \left( {\bf q}\right) $). The conductivity
(mobility) of particles is given by the solution of this set of equations, 
\begin{equation}
\sigma _{yy}^{}=\sigma _{zz}^{}=\sum_{j=1}^S\sigma _{yy}^{\left( j\right) }=-%
\frac{e^2L^2}{2\pi ^4\hbar \ell ^2}\sum_{j=1}^Sz_j^{1/2}\chi _j\left(
q_F^{\left( j\right) }\right) ,  \label{b2}
\end{equation}
and can be conveniently parametrized as 
\begin{equation}
\sigma _{yy}^{\left( j\right) }=\sigma _{zz}^{\left( j\right) }=\frac{e^2L^2%
}{\pi ^4\hbar \ell ^2}\Phi \left( z,\frac LR\right)   \label{b3}
\end{equation}
The functions $\Phi \left( z\right) $ for four different values of $R/L$ are
plotted in Fig. 1 ($R/L=0.05$) and Fig. 2 ($R/L=1;3;5$). The singular points
correspond to change in values of $S$ from $1$ to $2$ to $3$ to $4...$ at $%
z=3$, $13$, $34...$ 
\begin{figure}
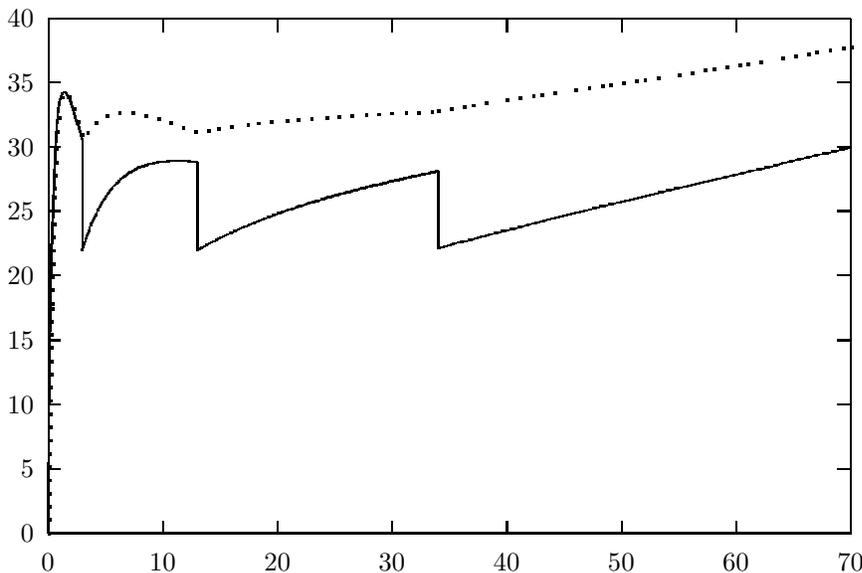

\centerline{\input r_difig1.tex}
\caption [F1]{Function $\Phi(x)$, Eq.(\ref{b3}), for $R/L=0.05$ (solid line); dashed line - the same function calculated without interlevel transitions}
\label{fig.1}
\end{figure}
\begin{figure}
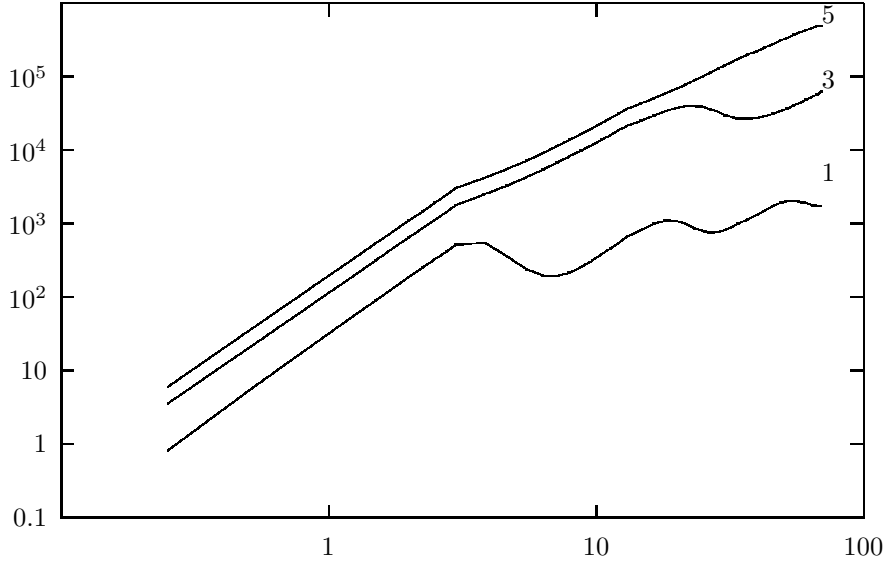

\centerline{\input r_difig2.tex}
\caption [F2]{Function $\Phi(x)$, Eq.(\ref{b3}), for $R/L=1;3;5$}
\label{fig.2}
\end{figure}

Another way of parameterization of equations, similar to the one used in 
{\rm I}, is based on definition 
\[
\frac{2\pi ^2\sqrt{z_jz_{j^{\prime }}}R^2}{L^2}=\frac{4\pi \sqrt{%
z_jz_{j^{\prime }}}NR^2}z=\frac{8\pi ^2\sqrt{z_jz_{j^{\prime }}}}z\left( 
\frac R\lambda \right) ^2, 
\]
where the effective particle wavelength $\lambda ^2=2\pi /N$. This equation
redefines the function $\Phi \left( z,L/R\right) $ as $\widetilde{\Phi }%
\left( z,\lambda /R\right) =\Phi \left( z,\sqrt{z}\lambda /2R\right) $.

Dramatic difference in shapes of the curves in Figs. 1 and 2 for small and
large values of $R/L$ is explained by the role of interlevel transitions. If
one simply neglects the interlevel transitions (the off-diagonal terms) in
the collision integral $\left( \text{\ref{a11}}\right) $, then the set of
transport equations $\left( \text{\ref{b1}}\right) $ will decouple into $S$
independent equations. It is fairly obvious that in this approximation the
conductivity $\left( \text{\ref{b2}}\right) $ is almost always a monotonic
function of $z=2NL^2/\pi $, though the critical values of $z$, which
correspond to the change in the number of occupied levels $S$, are still
responsible for the singularities (kinks) in the curves. Therefore, the
saw-like nature of the curves is caused by the interlevel transitions
exclusively.

For comparison, Fig. 1 (dashed line) and Fig. 3 give the function $\Phi
\left( z\right) $ for on-level scattering exclusively, {\it i.e.}, when all
the interlevel off-diagonal terms in the collision integral $\left( \text{%
\ref{a11}}\right) $ are artificially neglected. Obviously, the curves with
and without transitions always coincide as far as $z\leq 3$ and there is
only one occupied level. The differences show up only at $z>3$. 
\begin{figure}
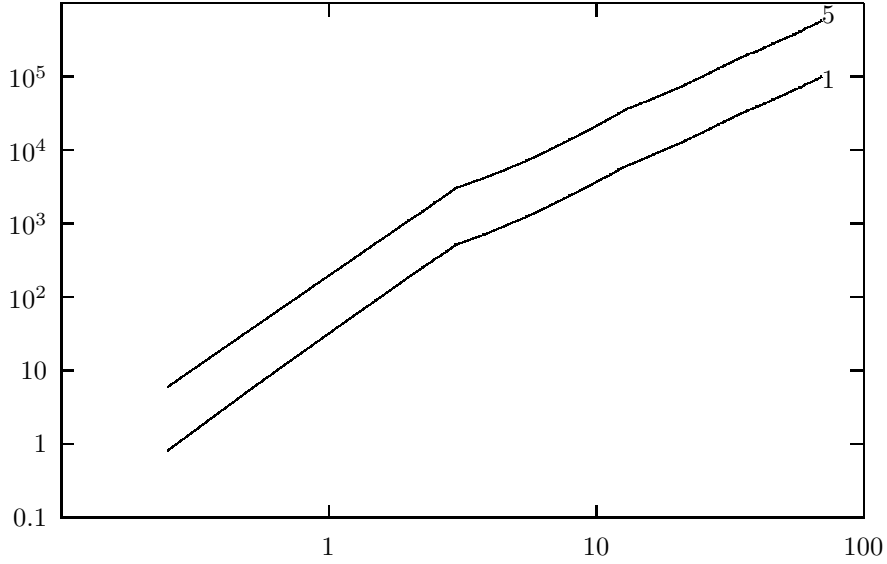

\centerline{\input r_difig3.tex}
\caption [F3]{Function $\Phi(x)$, Eq.(\ref{b3}), for $R/L=1;5$ caculated without interlevel transitions}
\label{fig.3}
\end{figure}

On the other hand, the importance of interlevel transitions is characterized
by the parameter $NR^2\sim zR^2/L^2$. Since $z_{j^{\prime
}}-z_j=j^2-j^{\prime 2}$, the exponent 
\[
\exp \left[ -\pi ^2\left( \sqrt{z_j}-\sqrt{z_{j^{\prime }}}\right)
^2R^2/2L^2\right] \equiv \exp \left[ -\pi \left( \sqrt{N_jR^2}-\sqrt{%
N_{j^{\prime }}R^2}\right) ^2\right] 
\]
always makes interlevel transitions to remote levels $\left| j-j^{\prime
}\right| \gg L/R$ negligible. These exponents show that the interlevel
transitions and the resulting mixing of adjacent levels are very important
only for not very populated levels with $2\pi ^2z_jR^2/L^2\ll 1$. Thus, the
contribution of interlevel transitions is very noticeable only for
relatively small values of $R/L,$ and decreases exponentially with
increasing $R/L$. For this reason, the saw-like character of particles
mobility becomes less and less pronounced with increasing $R/L$. At $R/L=5$
the saw nearly completely disappears, and there is practically no difference
between the curves in Figs. 2 (exact calculation) and Fig. 3 (calculation
with artificially neglected interlevel transitions). Note, that the curves
calculated with and without transitions always coincide for a small number
of particles $z<3$ when only one level is occupied and the transitions are
impossible for energy reasons.

The same parameters, $2\pi ^2z_jR^2/L^2=4\pi N_jR^2$, determine the values
of the hypergeometric functions in $\left( \text{\ref{b1}}\right) $. At $%
x^2\ll 1$, $_1F_1\left( \left( 2n-1\right) /2,n,-x^2\right) \simeq 1$, while
in the opposite case $x^2\gg 1$, $_1F_1\left( \left( 2n-1\right)
/2,n,-x^2\right) \simeq \left( n-1\right) !/\sqrt{\pi }x^n.$ Therefore, at
large $2\pi ^2z_jR^2/L^2\gg 1$, one can not only neglect the interlevel
transitions, but also the hypergeometric functions with $n=3$ and $n=4$ in
the diagonal terms of the collision integrals $\left( \text{\ref{b1}}\right) 
$ in comparison with the one with $n=2$. Only under these conditions one can
justify a heuristic assumption made in {\rm I} and recover the result $%
\left( {\rm I}.53\right) $: 
\begin{eqnarray}
\sigma _{yy} &=&\sum_{j=1}^S\sigma _{yy}^{\left( j\right) }=\frac{e^2L^2}{%
\pi ^4\hbar \ell ^2}\Phi \left( z,\frac RL\right) ,  \label{a20} \\
\Phi \left( z,\frac RL\right) &=&\frac{L^2}{4R^2}\sum_1^{S\left( z\right) }%
\frac 1{j^4}\frac{z_j}{_1F_1\left( \frac 32,2,-2\pi ^2z_jR^2/L^2\right) }. 
\nonumber
\end{eqnarray}

For small correlation radii $NR^2\ll 1$, all the terms in $\left( \text{\ref
{b1}}\right) $ are of the same order, the hypergeometric function $%
_1F_1\left( \left( 2n-1\right) /2,n,-2\pi ^2z_jR^2/L^2\right) \sim
\,_1F_1\left( \left( 2n-1\right) /2,n,0\right) =1$, and Eqs.$\left( \text{%
\ref{b1}}\right) $ can be simplified as 
\begin{equation}
\frac{z_j^{1/2}L^2}{R^2}=-\frac 12\chi _j\left( 4j^4+6z_jj^2+\frac 52%
z_j^2\,\right) -2\chi _j\sum_{j^{\prime }}^{S\left( z\right) }\left(
1-\delta _{jj^{\prime }}\right) j^2j^{\prime 2}  \label{b4}
\end{equation}
Then 
\begin{equation}
\Phi \left( z,\frac RL\right) =\frac{L^2}{4R^2}\sum_{j=1}^{S\left( z\right) }%
\frac{\nu \left( z\right) -j^2}{\left( j^4+3z_jj^2/2+5z_j^2/8\,\right)
+S\left( S+1\right) \left( 2S+1\right) /6-j^2}  \label{b5}
\end{equation}
In the opposite limiting case when for all levels $z_jR^2/L^2\gg 1,$ the
interlevel transitions and higher-order hypergeometric functions can be
neglected, $_1F_1\left( \frac 32,2,-x^2\right) \rightarrow 1/\sqrt{\pi }x^3$%
, and 
\begin{equation}
\Phi \left( z,\frac RL\right) =\frac{\pi ^{7/2}R}{2^{1/2}L}%
\sum_{j=1}^{S\left( z\right) }\frac{\left( \nu \left( z\right) -j^2\right)
^{5/2}}{j^4}  \label{b6}
\end{equation}
Note that Eq.$\left( \text{\ref{b6}}\right) $ for large $NR^2$ can be
improved near the critical values of $z$ which correspond to changes in
number of occupied levels $S$. With the appearance of a new level $S$, the
number of particles on this level, $z_S$, and, therefore, $z_SR^2/L^2$ are
small even for large $R/L$, and the contribution of this level is $%
z_S/\,_1F_1\left( \frac 32,2,-2\pi ^2z_sR^2/L^2\right) \sim z_S$, and not $%
z_S\pi ^{1/2}\left( 2\pi ^2z_SR^2/L^2\right) ^{3/2}$ as it is implied by Eq.$%
\left( \text{\ref{b6}}\right) $. Only later, when the hypergeometric
function will become small, $2\pi ^2z_SR^2/L^2\gg 1$, the contribution of
this highest level will recover the form indicated by $\left( \text{\ref{b6}}%
\right) $.

The argument of the exponents and hypergeometric functions can be also
written as the ratio of the particle wavelength to the correlation radius of
surface inhomogeneities, $2\pi ^2z_jR^2/L^2\sim \left( R/\lambda _j\right)
^2 $. Therefore, as it was mentioned in the Introduction, the particle
wavelength serves as a natural scale for describing the correlations and
separates long-range from short-range correlations. In this sense, the
interlevel transitions are more important for the short-range correlations.

The diffusion coefficient is related to the mobility $\left( \text{\ref{a20}}%
\right) $ as 
\begin{equation}
D_{yy}=D_{zz}=-\sigma _{yy}/e^2\sum_j\int \frac{\partial n_j}{\partial
\epsilon }\frac{m\,d\epsilon }{\pi \hbar ^2}=\frac{\pi \hbar ^2}{e^2mS}%
\sigma _{yy}=\frac{L^2\hbar }{\pi ^3m\ell ^2S}\Phi \left( z,\frac RL\right) ,
\label{a21}
\end{equation}
while the mean free path along the channel is 
\begin{equation}
{\cal L}=\sigma <q>/e^2N=\left( 2\pi \right) ^{1/2}\hbar \sigma \left( \sum
N_j^2\right) ^{1/2}/e^2N^{3/2}  \label{a22}
\end{equation}

\section{Transport Along Films and Channels: High Temperatures}

At finite temperatures all the levels with different $j$ are populated. As a
result, the transport equation becomes an infinite set of coupled equations.
The chemical potential is the same for particles on all levels, 
\begin{equation}
\mu \left( N,T\right) =\frac 1{2m}\left( \frac{\pi j\hbar }L\right) ^2+\mu
_j\left( N_j,T\right) ,\ \epsilon _j\left( {\bf q}\right) =\frac 1{2m}\left(
\left( \frac{\pi j\hbar }L\right) ^2+q^2\right)  \label{a27}
\end{equation}
where $\mu _j$ is the chemical potential of a $2D$ system of $N_j$ fermions
on level $j$. If we are dealing with a dilute gas, then $\mu _j$ depends
only on the number of particles on the corresponding level $z_j=2N_jL^2/\pi $%
, 
\[
z_j=\vartheta _T\ln \left( 1+\exp \left( \frac{\mu _j}T\right) \right)
=\vartheta _T\ln \left( 1+\exp \left( \frac \mu T-\frac{j^2}{\vartheta _T}%
\right) \right) , 
\]
and 
\begin{equation}
\mu _j=T\,\ln \left( \exp \left( \frac{z_j}{\vartheta _T}\right) -1\right)
\label{a28}
\end{equation}
where 
\[
\vartheta _T=\frac{2mTL^2}{\pi ^2\hbar ^2} 
\]
describes the ratio of the temperature to the energy of zero-point
oscillations in the well of the width $L.$ This equation should be used to
express the chemical potential via the total number of particles $%
z=2NL^2/\pi $, 
\begin{equation}
z=\vartheta _T\sum_1^\infty \ln \left[ \exp \left( \frac \mu T-\frac{j^2}{%
\vartheta _T}\right) +1\right]  \label{a29}
\end{equation}
The solution of this equation $\mu \left( z\right) $ at $T=0$ is given by
Eq. $\left( \text{\ref{a18}}\right) $.

In this Section we will calculate transport coefficients for particles with
Boltzmann distribution function. In the case of high-temperature Boltzmann
systems, 
\[
z\sim \vartheta _T\exp \left( \frac \mu T\right) \Theta ,\ \mu =T\,\ln
\left( \frac z{\vartheta _T\Theta }\right) ,\ \Theta \left( \vartheta
_T\right) \equiv \sum_{j=1}^\infty \exp \left( -\frac{j^2}{\vartheta _T}%
\right) 
\]
and the transport equation $\left( \text{\ref{a7}}\right) $ in dimensionless
variables $\chi _j$, 
\[
n_j\left( {\bf q}\right) =n_j^{\left( 0\right) }\left( q\right) \left( 1-%
\frac{FL^3}{\pi ^4T\ell ^2}\chi _j\cos \theta \right) 
\]
assumes the form 
\begin{eqnarray}
u^{1/2} &=&\frac 1{2\pi ^2L^2\ell ^2}\{\chi _j\left( q\right) \left[ \frac 14%
u^2\left( \eta _1\left( u,u\right) -\eta _0\left( u,u\right) \right)
+uj^2\left( \gamma _1\left( u,u\right) -\gamma _0\left( u,u\right) \right)
+j^4\left( \zeta _1\left( u,u\right) -\zeta _0\left( u,u\right) \right)
\right]  \nonumber \\
&&+\sum_{j^{\prime }\neq j}^{S\left( u,j\right) }j^2j^{\prime 2}\left[ \chi
_{j^{\prime }}\zeta _1\left( u,u_{jj^{\prime }}\right) -\chi _j\zeta
_0\left( u,u_{jj^{\prime }}\right) \right] \}  \label{a30}
\end{eqnarray}
while the mobility (conductivity) is 
\begin{equation}
\sigma _{yy}^{}=\sigma _{zz}^{}=\sum_{j=1}^\infty \sigma _{yy}^{\left(
j\right) }=-\frac{e^2L^2}{2\pi ^4\hbar \ell ^2}\frac z{\vartheta _T^2\Theta }%
\sum_{j=1}^\infty \exp \left( -\frac{j^2}{\vartheta _T}\right) \int
u^{1/2}\chi _j\left( q\right) \exp \left[ -\frac u{\vartheta _T}\right] \,du
\label{a31}
\end{equation}
Here 
\[
u=q^2\left( \frac L{\pi \hbar }\right) ^2,\ S\left( u,j\right) ={\rm Int\,}%
\left[ \left( u+j^2\right) ^{1/2}\right] ,\ u_{jj^{\prime }}=u+j^2-j^{\prime
2},\ 
\]
and $\zeta _i$, $\eta _i$, and $\gamma _{i\text{ }}$are the corresponding
angular Fourier harmonics of the functions 
\begin{eqnarray}
\zeta \left( {\bf q-q}^{\prime }\right) &=&\zeta \left( q^2+q^{\prime
2}-2qq^{\prime }\cos \varphi \right) ,\ \eta \left( {\bf q-q}^{\prime
}\right) =\zeta \left( {\bf q-q}^{\prime }\right) \left[ 1-\cos \varphi
\right] ^2,  \label{a32} \\
\gamma \left( {\bf q-q}^{\prime }\right) &=&\zeta \left( {\bf q-q}^{\prime
}\right) \left[ 1-\cos \varphi \right]  \nonumber
\end{eqnarray}
over the angle $\varphi $. In essence, the variable $u=\left( qL/\pi \hbar
\right) ^2$ plays the same role as the Fermi momenta $z_j=\left( q_F^{\left(
j\right) }L/\pi \hbar \right) ^2$ for degenerate systems in the previous
Section.

In the Gaussian case $\left( \text{\ref{a2}}\right) $, integration in $%
\left( \text{\ref{a30}}\right) $ leads, of course, to the same set of
equations $\left( \text{\ref{b1}}\right) $ with the only difference that $z_i
$ should be substituted by $u$. The situation is again non-analytic since
the summation in $\left( \text{\ref{a32}}\right) $ for off-diagonal
transitions over $j^{\prime }$ should be performed up to the value $S\left(
u,j\right) $ which is not only different for each $j$, {\it i.e.,} for each
equation, but also depends on momentum $q$ and exhibits step-like jumps at
certain values of $u=q^2\left( L/\pi \hbar \right) ^2$. However, this
non-analyticity manifests itself more noticeably in the integrands $\left( 
\text{\ref{a31}}\right) $ rather than in transport coefficients themselves
which are fairly smooth. Finally, the conductivity (mobility) is equal to 
\begin{eqnarray}
\sigma _{yy} &=&\sum_{j=1}^\infty \sigma _{yy}^{\left( j\right) }=\frac{%
e^2NL^4}{\pi ^5\hbar \ell ^2}\Pi \left( \vartheta _T,\frac RL\right) ,
\label{a33} \\
\Pi \left( x,y\right)  &=&\frac 1{x^2\Theta \left( x\right) y^2}%
\sum_{j=1}^\infty \int \chi _j\left( u\right) \exp \left( -\frac{j^2+u^2}x%
\right) \,du  \nonumber
\end{eqnarray}
Function $\Pi \left( \vartheta _T\right) $ is plotted in Fig. 4 ($R/L=0.05$)
and Fig. 5 ($R/L=0.5;\ 1$). 
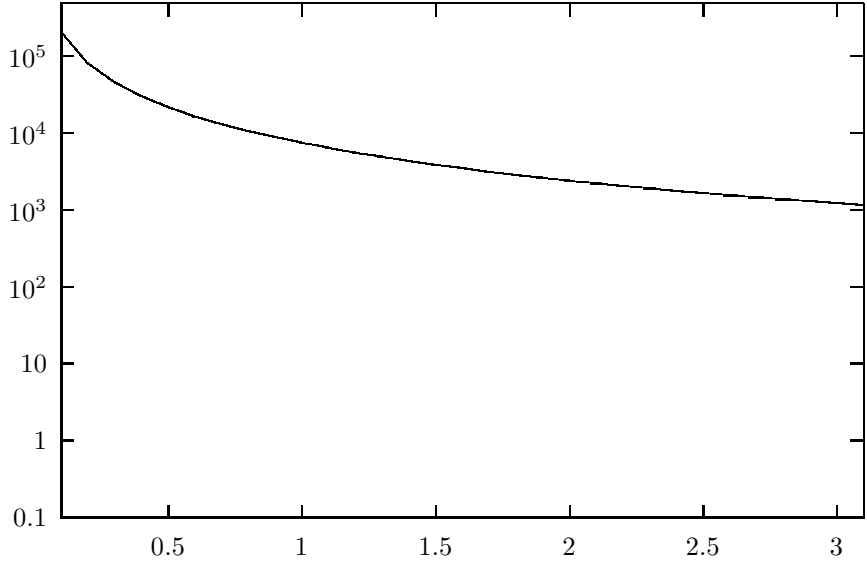
\begin{figure}
\centerline{
\setlength{\unitlength}{0.240900pt}
\ifx\plotpoint\undefined\newsavebox{\plotpoint}\fi
\sbox{\plotpoint}{\rule[-0.200pt]{0.400pt}{0.400pt}}%
\begin{picture}(1500,900)(0,0)
\font\gnuplot=cmr10 at 10pt
\gnuplot
\sbox{\plotpoint}{\rule[-0.200pt]{0.400pt}{0.400pt}}%
\put(176.0,68.0){\rule[-0.200pt]{4.818pt}{0.400pt}}
\put(154,68){\makebox(0,0)[r]{0.1}}
\put(1416.0,68.0){\rule[-0.200pt]{4.818pt}{0.400pt}}
\put(176.0,189.0){\rule[-0.200pt]{4.818pt}{0.400pt}}
\put(154,189){\makebox(0,0)[r]{1}}
\put(1416.0,189.0){\rule[-0.200pt]{4.818pt}{0.400pt}}
\put(176.0,310.0){\rule[-0.200pt]{4.818pt}{0.400pt}}
\put(154,310){\makebox(0,0)[r]{10}}
\put(1416.0,310.0){\rule[-0.200pt]{4.818pt}{0.400pt}}
\put(176.0,430.0){\rule[-0.200pt]{4.818pt}{0.400pt}}
\put(154,430){\makebox(0,0)[r]{$10^2$}}
\put(1416.0,430.0){\rule[-0.200pt]{4.818pt}{0.400pt}}
\put(176.0,551.0){\rule[-0.200pt]{4.818pt}{0.400pt}}
\put(154,551){\makebox(0,0)[r]{$10^3$}}
\put(1416.0,551.0){\rule[-0.200pt]{4.818pt}{0.400pt}}
\put(176.0,672.0){\rule[-0.200pt]{4.818pt}{0.400pt}}
\put(154,672){\makebox(0,0)[r]{$10^4$}}
\put(1416.0,672.0){\rule[-0.200pt]{4.818pt}{0.400pt}}
\put(176.0,793.0){\rule[-0.200pt]{4.818pt}{0.400pt}}
\put(154,793){\makebox(0,0)[r]{$10^5$}}
\put(1416.0,793.0){\rule[-0.200pt]{4.818pt}{0.400pt}}
\put(344.0,68.0){\rule[-0.200pt]{0.400pt}{4.818pt}}
\put(344,23){\makebox(0,0){0.5}}
\put(344.0,857.0){\rule[-0.200pt]{0.400pt}{4.818pt}}
\put(554.0,68.0){\rule[-0.200pt]{0.400pt}{4.818pt}}
\put(554,23){\makebox(0,0){1}}
\put(554.0,857.0){\rule[-0.200pt]{0.400pt}{4.818pt}}
\put(764.0,68.0){\rule[-0.200pt]{0.400pt}{4.818pt}}
\put(764,23){\makebox(0,0){1.5}}
\put(764.0,857.0){\rule[-0.200pt]{0.400pt}{4.818pt}}
\put(974.0,68.0){\rule[-0.200pt]{0.400pt}{4.818pt}}
\put(974,23){\makebox(0,0){2}}
\put(974.0,857.0){\rule[-0.200pt]{0.400pt}{4.818pt}}
\put(1184.0,68.0){\rule[-0.200pt]{0.400pt}{4.818pt}}
\put(1184,23){\makebox(0,0){2.5}}
\put(1184.0,857.0){\rule[-0.200pt]{0.400pt}{4.818pt}}
\put(1394.0,68.0){\rule[-0.200pt]{0.400pt}{4.818pt}}
\put(1394,23){\makebox(0,0){3}}
\put(1394.0,857.0){\rule[-0.200pt]{0.400pt}{4.818pt}}
\put(176.0,68.0){\rule[-0.200pt]{303.534pt}{0.400pt}}
\put(1436.0,68.0){\rule[-0.200pt]{0.400pt}{194.888pt}}
\put(176.0,877.0){\rule[-0.200pt]{303.534pt}{0.400pt}}
\put(176.0,68.0){\rule[-0.200pt]{0.400pt}{194.888pt}}
\put(176,832){\usebox{\plotpoint}}
\multiput(176.58,829.61)(0.498,-0.595){81}{\rule{0.120pt}{0.576pt}}
\multiput(175.17,830.80)(42.000,-48.804){2}{\rule{0.400pt}{0.288pt}}
\multiput(218.00,780.92)(0.701,-0.497){57}{\rule{0.660pt}{0.120pt}}
\multiput(218.00,781.17)(40.630,-30.000){2}{\rule{0.330pt}{0.400pt}}
\multiput(260.00,750.92)(0.960,-0.496){41}{\rule{0.864pt}{0.120pt}}
\multiput(260.00,751.17)(40.207,-22.000){2}{\rule{0.432pt}{0.400pt}}
\multiput(302.00,728.92)(1.249,-0.495){31}{\rule{1.088pt}{0.119pt}}
\multiput(302.00,729.17)(39.741,-17.000){2}{\rule{0.544pt}{0.400pt}}
\multiput(344.00,711.92)(1.421,-0.494){27}{\rule{1.220pt}{0.119pt}}
\multiput(344.00,712.17)(39.468,-15.000){2}{\rule{0.610pt}{0.400pt}}
\multiput(386.00,696.92)(1.789,-0.492){21}{\rule{1.500pt}{0.119pt}}
\multiput(386.00,697.17)(38.887,-12.000){2}{\rule{0.750pt}{0.400pt}}
\multiput(428.00,684.92)(1.958,-0.492){19}{\rule{1.627pt}{0.118pt}}
\multiput(428.00,685.17)(38.623,-11.000){2}{\rule{0.814pt}{0.400pt}}
\multiput(470.00,673.93)(2.417,-0.489){15}{\rule{1.967pt}{0.118pt}}
\multiput(470.00,674.17)(37.918,-9.000){2}{\rule{0.983pt}{0.400pt}}
\multiput(512.00,664.93)(2.417,-0.489){15}{\rule{1.967pt}{0.118pt}}
\multiput(512.00,665.17)(37.918,-9.000){2}{\rule{0.983pt}{0.400pt}}
\multiput(554.00,655.93)(2.739,-0.488){13}{\rule{2.200pt}{0.117pt}}
\multiput(554.00,656.17)(37.434,-8.000){2}{\rule{1.100pt}{0.400pt}}
\multiput(596.00,647.93)(2.739,-0.488){13}{\rule{2.200pt}{0.117pt}}
\multiput(596.00,648.17)(37.434,-8.000){2}{\rule{1.100pt}{0.400pt}}
\multiput(638.00,639.93)(3.745,-0.482){9}{\rule{2.900pt}{0.116pt}}
\multiput(638.00,640.17)(35.981,-6.000){2}{\rule{1.450pt}{0.400pt}}
\multiput(680.00,633.93)(3.162,-0.485){11}{\rule{2.500pt}{0.117pt}}
\multiput(680.00,634.17)(36.811,-7.000){2}{\rule{1.250pt}{0.400pt}}
\multiput(722.00,626.93)(3.745,-0.482){9}{\rule{2.900pt}{0.116pt}}
\multiput(722.00,627.17)(35.981,-6.000){2}{\rule{1.450pt}{0.400pt}}
\multiput(764.00,620.93)(4.606,-0.477){7}{\rule{3.460pt}{0.115pt}}
\multiput(764.00,621.17)(34.819,-5.000){2}{\rule{1.730pt}{0.400pt}}
\multiput(806.00,615.93)(3.745,-0.482){9}{\rule{2.900pt}{0.116pt}}
\multiput(806.00,616.17)(35.981,-6.000){2}{\rule{1.450pt}{0.400pt}}
\multiput(848.00,609.93)(4.606,-0.477){7}{\rule{3.460pt}{0.115pt}}
\multiput(848.00,610.17)(34.819,-5.000){2}{\rule{1.730pt}{0.400pt}}
\multiput(890.00,604.94)(6.038,-0.468){5}{\rule{4.300pt}{0.113pt}}
\multiput(890.00,605.17)(33.075,-4.000){2}{\rule{2.150pt}{0.400pt}}
\multiput(932.00,600.93)(4.606,-0.477){7}{\rule{3.460pt}{0.115pt}}
\multiput(932.00,601.17)(34.819,-5.000){2}{\rule{1.730pt}{0.400pt}}
\multiput(974.00,595.94)(6.038,-0.468){5}{\rule{4.300pt}{0.113pt}}
\multiput(974.00,596.17)(33.075,-4.000){2}{\rule{2.150pt}{0.400pt}}
\multiput(1016.00,591.94)(6.038,-0.468){5}{\rule{4.300pt}{0.113pt}}
\multiput(1016.00,592.17)(33.075,-4.000){2}{\rule{2.150pt}{0.400pt}}
\multiput(1058.00,587.94)(6.038,-0.468){5}{\rule{4.300pt}{0.113pt}}
\multiput(1058.00,588.17)(33.075,-4.000){2}{\rule{2.150pt}{0.400pt}}
\multiput(1100.00,583.94)(6.038,-0.468){5}{\rule{4.300pt}{0.113pt}}
\multiput(1100.00,584.17)(33.075,-4.000){2}{\rule{2.150pt}{0.400pt}}
\multiput(1142.00,579.95)(9.169,-0.447){3}{\rule{5.700pt}{0.108pt}}
\multiput(1142.00,580.17)(30.169,-3.000){2}{\rule{2.850pt}{0.400pt}}
\multiput(1184.00,576.94)(6.038,-0.468){5}{\rule{4.300pt}{0.113pt}}
\multiput(1184.00,577.17)(33.075,-4.000){2}{\rule{2.150pt}{0.400pt}}
\multiput(1226.00,572.95)(9.169,-0.447){3}{\rule{5.700pt}{0.108pt}}
\multiput(1226.00,573.17)(30.169,-3.000){2}{\rule{2.850pt}{0.400pt}}
\multiput(1268.00,569.95)(9.169,-0.447){3}{\rule{5.700pt}{0.108pt}}
\multiput(1268.00,570.17)(30.169,-3.000){2}{\rule{2.850pt}{0.400pt}}
\multiput(1310.00,566.95)(9.169,-0.447){3}{\rule{5.700pt}{0.108pt}}
\multiput(1310.00,567.17)(30.169,-3.000){2}{\rule{2.850pt}{0.400pt}}
\multiput(1352.00,563.95)(9.169,-0.447){3}{\rule{5.700pt}{0.108pt}}
\multiput(1352.00,564.17)(30.169,-3.000){2}{\rule{2.850pt}{0.400pt}}
\multiput(1394.00,560.95)(9.169,-0.447){3}{\rule{5.700pt}{0.108pt}}
\multiput(1394.00,561.17)(30.169,-3.000){2}{\rule{2.850pt}{0.400pt}}
\end{picture}}
\caption [F4]{Function $\Pi(x)$, Eq.(\ref{a33}), for $R/L=0.05$}
\label{fig.4}
\end{figure}
\begin{figure}
\centerline{
\setlength{\unitlength}{0.240900pt}
\ifx\plotpoint\undefined\newsavebox{\plotpoint}\fi
\sbox{\plotpoint}{\rule[-0.200pt]{0.400pt}{0.400pt}}%
\begin{picture}(1500,900)(0,0)
\font\gnuplot=cmr10 at 10pt
\gnuplot
\sbox{\plotpoint}{\rule[-0.200pt]{0.400pt}{0.400pt}}%
\put(176.0,68.0){\rule[-0.200pt]{303.534pt}{0.400pt}}
\put(176.0,68.0){\rule[-0.200pt]{4.818pt}{0.400pt}}
\put(154,68){\makebox(0,0)[r]{0}}
\put(1416.0,68.0){\rule[-0.200pt]{4.818pt}{0.400pt}}
\put(176.0,215.0){\rule[-0.200pt]{4.818pt}{0.400pt}}
\put(154,215){\makebox(0,0)[r]{10}}
\put(1416.0,215.0){\rule[-0.200pt]{4.818pt}{0.400pt}}
\put(176.0,362.0){\rule[-0.200pt]{4.818pt}{0.400pt}}
\put(154,362){\makebox(0,0)[r]{20}}
\put(1416.0,362.0){\rule[-0.200pt]{4.818pt}{0.400pt}}
\put(176.0,509.0){\rule[-0.200pt]{4.818pt}{0.400pt}}
\put(154,509){\makebox(0,0)[r]{30}}
\put(1416.0,509.0){\rule[-0.200pt]{4.818pt}{0.400pt}}
\put(176.0,656.0){\rule[-0.200pt]{4.818pt}{0.400pt}}
\put(154,656){\makebox(0,0)[r]{40}}
\put(1416.0,656.0){\rule[-0.200pt]{4.818pt}{0.400pt}}
\put(176.0,803.0){\rule[-0.200pt]{4.818pt}{0.400pt}}
\put(154,803){\makebox(0,0)[r]{50}}
\put(1416.0,803.0){\rule[-0.200pt]{4.818pt}{0.400pt}}
\put(344.0,68.0){\rule[-0.200pt]{0.400pt}{4.818pt}}
\put(344,23){\makebox(0,0){0.5}}
\put(344.0,857.0){\rule[-0.200pt]{0.400pt}{4.818pt}}
\put(554.0,68.0){\rule[-0.200pt]{0.400pt}{4.818pt}}
\put(554,23){\makebox(0,0){1}}
\put(554.0,857.0){\rule[-0.200pt]{0.400pt}{4.818pt}}
\put(764.0,68.0){\rule[-0.200pt]{0.400pt}{4.818pt}}
\put(764,23){\makebox(0,0){1.5}}
\put(764.0,857.0){\rule[-0.200pt]{0.400pt}{4.818pt}}
\put(974.0,68.0){\rule[-0.200pt]{0.400pt}{4.818pt}}
\put(974,23){\makebox(0,0){2}}
\put(974.0,857.0){\rule[-0.200pt]{0.400pt}{4.818pt}}
\put(1184.0,68.0){\rule[-0.200pt]{0.400pt}{4.818pt}}
\put(1184,23){\makebox(0,0){2.5}}
\put(1184.0,857.0){\rule[-0.200pt]{0.400pt}{4.818pt}}
\put(1394.0,68.0){\rule[-0.200pt]{0.400pt}{4.818pt}}
\put(1394,23){\makebox(0,0){3}}
\put(1394.0,857.0){\rule[-0.200pt]{0.400pt}{4.818pt}}
\put(176.0,68.0){\rule[-0.200pt]{303.534pt}{0.400pt}}
\put(1436.0,68.0){\rule[-0.200pt]{0.400pt}{194.888pt}}
\put(176.0,877.0){\rule[-0.200pt]{303.534pt}{0.400pt}}
\put(1352,392){\makebox(0,0)[l]{1}}
\put(1352,509){\makebox(0,0)[l]{0.5}}
\put(176.0,68.0){\rule[-0.200pt]{0.400pt}{194.888pt}}
\sbox{\plotpoint}{\rule[-0.500pt]{1.000pt}{1.000pt}}%
\put(176,827){\usebox{\plotpoint}}
\multiput(176,827)(2.530,-20.601){17}{\usebox{\plotpoint}}
\multiput(218,485)(7.463,-19.367){6}{\usebox{\plotpoint}}
\multiput(260,376)(13.194,-16.022){3}{\usebox{\plotpoint}}
\multiput(302,325)(18.386,-9.631){2}{\usebox{\plotpoint}}
\multiput(344,303)(20.662,-1.968){2}{\usebox{\plotpoint}}
\multiput(386,299)(20.191,4.807){2}{\usebox{\plotpoint}}
\multiput(428,309)(19.077,8.176){3}{\usebox{\plotpoint}}
\multiput(470,327)(18.386,9.631){2}{\usebox{\plotpoint}}
\multiput(512,349)(18.021,10.298){2}{\usebox{\plotpoint}}
\multiput(554,373)(18.021,10.298){2}{\usebox{\plotpoint}}
\multiput(596,397)(18.564,9.282){3}{\usebox{\plotpoint}}
\multiput(638,418)(18.739,8.923){2}{\usebox{\plotpoint}}
\multiput(680,438)(19.396,7.389){2}{\usebox{\plotpoint}}
\multiput(722,454)(19.690,6.563){2}{\usebox{\plotpoint}}
\multiput(764,468)(20.078,5.259){2}{\usebox{\plotpoint}}
\multiput(806,479)(20.295,4.349){2}{\usebox{\plotpoint}}
\multiput(848,488)(20.547,2.935){2}{\usebox{\plotpoint}}
\multiput(890,494)(20.610,2.454){3}{\usebox{\plotpoint}}
\multiput(932,499)(20.732,0.987){2}{\usebox{\plotpoint}}
\multiput(974,501)(20.732,0.987){2}{\usebox{\plotpoint}}
\multiput(1016,503)(20.750,-0.494){2}{\usebox{\plotpoint}}
\multiput(1058,502)(20.750,-0.494){2}{\usebox{\plotpoint}}
\multiput(1100,501)(20.732,-0.987){2}{\usebox{\plotpoint}}
\multiput(1142,499)(20.703,-1.479){2}{\usebox{\plotpoint}}
\multiput(1184,496)(20.662,-1.968){2}{\usebox{\plotpoint}}
\multiput(1226,492)(20.662,-1.968){2}{\usebox{\plotpoint}}
\multiput(1268,488)(20.610,-2.454){2}{\usebox{\plotpoint}}
\multiput(1310,483)(20.610,-2.454){2}{\usebox{\plotpoint}}
\multiput(1352,478)(20.547,-2.935){2}{\usebox{\plotpoint}}
\multiput(1394,472)(20.610,-2.454){2}{\usebox{\plotpoint}}
\put(1436,467){\usebox{\plotpoint}}
\sbox{\plotpoint}{\rule[-0.200pt]{0.400pt}{0.400pt}}%
\put(176,453){\usebox{\plotpoint}}
\multiput(176.58,448.67)(0.498,-1.183){81}{\rule{0.120pt}{1.043pt}}
\multiput(175.17,450.84)(42.000,-96.835){2}{\rule{0.400pt}{0.521pt}}
\multiput(218.58,351.73)(0.498,-0.559){81}{\rule{0.120pt}{0.548pt}}
\multiput(217.17,352.86)(42.000,-45.863){2}{\rule{0.400pt}{0.274pt}}
\multiput(260.00,305.92)(0.810,-0.497){49}{\rule{0.746pt}{0.120pt}}
\multiput(260.00,306.17)(40.451,-26.000){2}{\rule{0.373pt}{0.400pt}}
\multiput(302.00,279.92)(1.525,-0.494){25}{\rule{1.300pt}{0.119pt}}
\multiput(302.00,280.17)(39.302,-14.000){2}{\rule{0.650pt}{0.400pt}}
\multiput(344.00,265.94)(6.038,-0.468){5}{\rule{4.300pt}{0.113pt}}
\multiput(344.00,266.17)(33.075,-4.000){2}{\rule{2.150pt}{0.400pt}}
\multiput(386.00,263.60)(6.038,0.468){5}{\rule{4.300pt}{0.113pt}}
\multiput(386.00,262.17)(33.075,4.000){2}{\rule{2.150pt}{0.400pt}}
\multiput(428.00,267.58)(2.163,0.491){17}{\rule{1.780pt}{0.118pt}}
\multiput(428.00,266.17)(38.306,10.000){2}{\rule{0.890pt}{0.400pt}}
\multiput(470.00,277.58)(1.789,0.492){21}{\rule{1.500pt}{0.119pt}}
\multiput(470.00,276.17)(38.887,12.000){2}{\rule{0.750pt}{0.400pt}}
\multiput(512.00,289.58)(1.646,0.493){23}{\rule{1.392pt}{0.119pt}}
\multiput(512.00,288.17)(39.110,13.000){2}{\rule{0.696pt}{0.400pt}}
\multiput(554.00,302.58)(1.646,0.493){23}{\rule{1.392pt}{0.119pt}}
\multiput(554.00,301.17)(39.110,13.000){2}{\rule{0.696pt}{0.400pt}}
\multiput(596.00,315.58)(1.789,0.492){21}{\rule{1.500pt}{0.119pt}}
\multiput(596.00,314.17)(38.887,12.000){2}{\rule{0.750pt}{0.400pt}}
\multiput(638.00,327.58)(1.958,0.492){19}{\rule{1.627pt}{0.118pt}}
\multiput(638.00,326.17)(38.623,11.000){2}{\rule{0.814pt}{0.400pt}}
\multiput(680.00,338.59)(2.417,0.489){15}{\rule{1.967pt}{0.118pt}}
\multiput(680.00,337.17)(37.918,9.000){2}{\rule{0.983pt}{0.400pt}}
\multiput(722.00,347.59)(2.739,0.488){13}{\rule{2.200pt}{0.117pt}}
\multiput(722.00,346.17)(37.434,8.000){2}{\rule{1.100pt}{0.400pt}}
\multiput(764.00,355.59)(3.745,0.482){9}{\rule{2.900pt}{0.116pt}}
\multiput(764.00,354.17)(35.981,6.000){2}{\rule{1.450pt}{0.400pt}}
\multiput(806.00,361.60)(6.038,0.468){5}{\rule{4.300pt}{0.113pt}}
\multiput(806.00,360.17)(33.075,4.000){2}{\rule{2.150pt}{0.400pt}}
\multiput(848.00,365.61)(9.169,0.447){3}{\rule{5.700pt}{0.108pt}}
\multiput(848.00,364.17)(30.169,3.000){2}{\rule{2.850pt}{0.400pt}}
\multiput(890.00,368.61)(9.169,0.447){3}{\rule{5.700pt}{0.108pt}}
\multiput(890.00,367.17)(30.169,3.000){2}{\rule{2.850pt}{0.400pt}}
\put(932,370.67){\rule{10.118pt}{0.400pt}}
\multiput(932.00,370.17)(21.000,1.000){2}{\rule{5.059pt}{0.400pt}}
\put(1016,370.67){\rule{10.118pt}{0.400pt}}
\multiput(1016.00,371.17)(21.000,-1.000){2}{\rule{5.059pt}{0.400pt}}
\put(1058,369.67){\rule{10.118pt}{0.400pt}}
\multiput(1058.00,370.17)(21.000,-1.000){2}{\rule{5.059pt}{0.400pt}}
\put(1100,368.17){\rule{8.500pt}{0.400pt}}
\multiput(1100.00,369.17)(24.358,-2.000){2}{\rule{4.250pt}{0.400pt}}
\multiput(1142.00,366.95)(9.169,-0.447){3}{\rule{5.700pt}{0.108pt}}
\multiput(1142.00,367.17)(30.169,-3.000){2}{\rule{2.850pt}{0.400pt}}
\put(1184,363.17){\rule{8.500pt}{0.400pt}}
\multiput(1184.00,364.17)(24.358,-2.000){2}{\rule{4.250pt}{0.400pt}}
\multiput(1226.00,361.95)(9.169,-0.447){3}{\rule{5.700pt}{0.108pt}}
\multiput(1226.00,362.17)(30.169,-3.000){2}{\rule{2.850pt}{0.400pt}}
\multiput(1268.00,358.94)(6.038,-0.468){5}{\rule{4.300pt}{0.113pt}}
\multiput(1268.00,359.17)(33.075,-4.000){2}{\rule{2.150pt}{0.400pt}}
\multiput(1310.00,354.95)(9.169,-0.447){3}{\rule{5.700pt}{0.108pt}}
\multiput(1310.00,355.17)(30.169,-3.000){2}{\rule{2.850pt}{0.400pt}}
\multiput(1352.00,351.94)(6.038,-0.468){5}{\rule{4.300pt}{0.113pt}}
\multiput(1352.00,352.17)(33.075,-4.000){2}{\rule{2.150pt}{0.400pt}}
\multiput(1394.00,347.94)(6.038,-0.468){5}{\rule{4.300pt}{0.113pt}}
\multiput(1394.00,348.17)(33.075,-4.000){2}{\rule{2.150pt}{0.400pt}}
\put(974.0,372.0){\rule[-0.200pt]{10.118pt}{0.400pt}}
\end{picture}}
\caption [F5]{Function $\Pi(x)$, Eq.(\ref{a33}), for $R/L= 0.5; 1$}
\label{fig.5}
\end{figure}
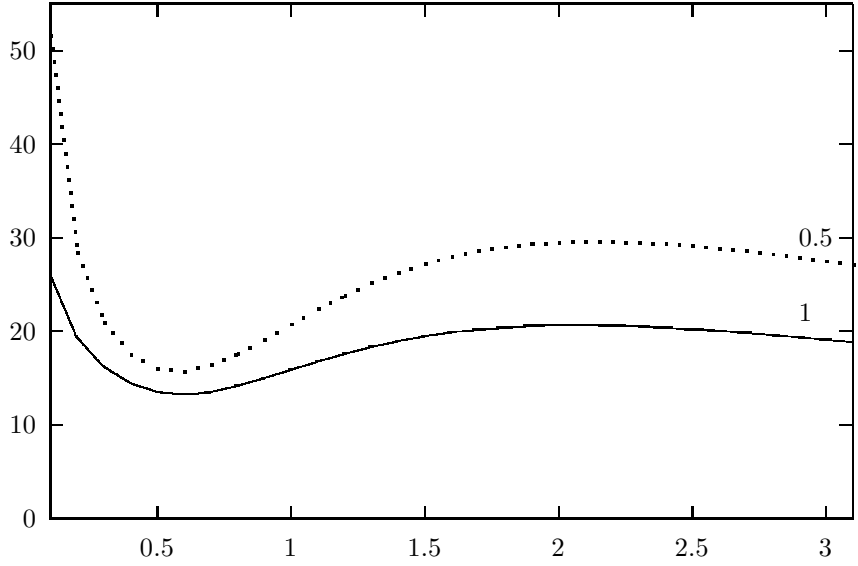

In the Boltzmann temperature range, the diffusion coefficient can be
expressed via mobility as 
\begin{equation}
D_{yy,zz}=-\sigma _{yy}/e^2\sum_j\int \frac{\partial n_j}{\partial \epsilon }%
\frac{m\,d\epsilon }{\pi \hbar ^2}=\frac{T\sigma _{yy}}{e^2N}=\frac{TL^4}{%
\pi ^5\hbar \ell ^2}\Pi \left( \vartheta _T,\frac RL\right) ,  \label{zz1}
\end{equation}
while the mean free path 
\begin{equation}
{\cal L}=\sigma <q>/e^2N=\frac{\left( mT\right) ^{1/2}L^4}{\pi ^5\hbar \ell
^2}\Pi \left( \vartheta _T,\frac RL\right)  \label{zz2}
\end{equation}

The difference between the functions $\Pi \left( x\right) $ in Figs.4,5 by
several orders of magnitude is not surprising. Since $x=\vartheta _T\sim
\left( L/\lambda \right) ^2$ ($\lambda $ is the particle wavelength), Fig.4
is potted in the region $L\sim \lambda $. On the other hand, $y=R/L=0.05$ is
rather small meaning that $R/\lambda \ll 1$. As it was explained in {\rm I}
(and is confirmed by the present calculation), condition $R/\lambda \ll 1$
corresponds to nearly specular quantum reflection, and, therefore, to large
particle mean free paths. Thus the large values of $\Pi \left( x\right) $ in
Fig.4. In Fig.5, $y=R/L\sim R/\lambda \sim 1$. This case corresponds to the
most effective scattering of particles by surface inhomogeneities and to the
smallest values of the mean free path.

\section{Summary and Discussion}

In summary, we calculated mobility and diffusion coefficients for ballistic
particles in ultra-narrow channels and films with random rough boundaries in
conditions when the motion of particles across the films is quantized. We
obtained explicit expressions for transport coefficients via the correlation
function of surface inhomogeneities. The most important consequence of a
discrete character of the particle spectrum for the motion across the film
is the non-analytic low-temperature dependence of the transport coefficients
on the film thickness and the density of particles.

The form of this non-analyticity strongly depends on the correlation radius
of surface inhomogeneities $R$. In the case of short-range correlations of
surface inhomogeneities, the low-temperature dependence of transport
coefficients on particle density and film thickness has a pronounced
saw-like structure. The saw teeth become smaller and the saw-like structure
gradually disappears with increasing correlation radius. Finally, for
long-range correlations one gets only not very well pronounced kinks,
instead of the saw teeth, at critical values of density and/or thickness at
which the number of occupied levels changes by one.

Though both the amplitude an the correlation radius of surface
inhomogeneities affect the particle scattering by the walls, the dependence
of transport coefficients on the amplitude of the surface inhomogeneities $%
\ell $, in contrast to their dependence on the correlation radius $R$, is
quite trivial, and reduces to a multiplicative factor $1/\ell ^2.$

In general, the non-analytic nature of the curves is explained by the
singularities in low-temperature distribution of degenerate fermions over a
system of discrete energy levels. However, the sharp discontinuities on the
saw-like curves for transport coefficients are caused not by the
singularities in the density of state, but mostly by the interlevel
transitions caused by the scattering on random rough walls. The beginning of
the occupation of a new level leads to two transport effects: to the direct
transport contribution of the particles from this new level, and to the
opening of new scattering channels for particles on all already occupied
levels (interlevel transitions to and from the new level). The first effect
is proportional to the number of particles on the new level and is small.
For this reason the non-analyticity of the transport coefficients reduces,
in the absence of interlevel transitions, to a series of kinks corresponding
to the appearance of the new levels. On the other hand, the opening of new
scattering channels with the interlevel transitions to and from newly
occupied levels affects particles on {\em all} already occupied levels thus
increasing dramatically the total effective scattering cross-section in a
step-like manner. If one artificially freezes these transitions, the
transport curves will exhibit kinks rather than the saw teeth. Not
surprisingly, the contribution of interlevel transitions depends
exponentially on the ratio of the particle wavelength to the correlation
radius of the surface inhomogeneities, and decreases rapidly with increasing
correlation radius of surface roughness ({\it i.e., }with flattening of
surface inhomogeneities).

Note, that the parameterization of transport parameters in this paper is
slightly different from {\rm I}. In the case of the mean free path it is,
probably, better to use, instead of $\left( \text{\ref{a22}}\right) ,\left( 
\text{\ref{zz2}}\right) $, the parameterization in the form {\rm I} 
\[
{\cal L\sim }\frac{L^2R}{\ell ^2}f\left( R/\lambda \right) 
\]
with the minimum at $R\sim \lambda $. The transformation of the results to
this form is fairly obvious in both degenerate and Boltzmann regions.

\section{ACKNOWLEDGMENTS.}

This work was supported by NSF grant DMR-9412769.

\section{Appendix. Classical and Semi-Classical Motion Across the Channels}

In the classical limit, when the distance between levels with different $j$
becomes negligible, the above results should coincide with the results of
classical calculations in Ref. {\rm I}. The transition to the classical
limit corresponds either to thick films and/or to the states with large
quantum numbers, $j\gg 1$, when the interlevel transitions are accompanied
by relatively small changes of the quantum number, $1\sim \delta j\ll j$.
The coordinate transformation $\left( \text{\ref{a1}}\right) $ and the
effective Hamiltonian $\left( \text{\ref{a5}}\right) $ are, obviously, the
same in classical and quantum cases. The matrix elements $\left( \text{\ref
{a10}}\right) $ of the effective bulk distortion $\left( \text{\ref{a5}}%
\right) $ in the classical limit correspond exactly to the classical matrix
elements in {\rm I}. However, the classical analog of the squares of these
matrix elements, and, therefore, the collision integral $\left( \text{\ref
{a11}}\right) $ are slightly different from Eqs.$\left( {\rm I}.17\right)
,\left( {\rm I}.18\right) ,\left( {\rm I.}20\right) $. The reason is that in 
{\rm I} we used not the best approximation for the following combination of
the $\delta $-functions: 
\begin{equation}
\delta ^{\prime }\left( {\bf p}_x-{\bf p}_x^{\prime }\right) \delta \left( 
{\bf p}_x-{\bf p}_x^{\prime }\right) =-\frac 12\left[ \delta ^2\left( {\bf p}%
_x-{\bf p}_x^{\prime }\right) \right] ^{\prime }\simeq -\frac L{2\hbar }%
\delta ^{\prime }\left( {\bf p}_x-{\bf p}_x^{\prime }\right)  \label{ap1}
\end{equation}

In classical calculations, quadratic expressions similar to $\left( \text{%
\ref{ap1}}\right) $ are not very well defined. More accurate expressions can
be obtained either by working with bell-shaped functions instead of $\delta $%
-functions ({\it i.e., }by introducing some small dissipation in the
equations), or by looking at the discrete (quantum) limit when all the
combinations of the Kronecker symbols $\delta _{ik}$ are well defined.
Careful analysis of the classical limit for the square of the matrix
elements $\left( \text{\ref{a10}}\right) $ demonstrates that in this
particular case the term in the {\it r.h.s. }of Eq.$\left( \text{\ref{ap1}}%
\right) $ should be written as $-\frac 1{2p_x}\delta ^{\prime }\left( {\bf p}%
_x-{\bf p}_x^{\prime }\right) $ rather than $-\frac L{2\hbar }\delta
^{\prime }\left( {\bf p}_x-{\bf p}_x^{\prime }\right) $ and neglected in the
classical limit $p_x\gg \hbar /L$ ({\it i.e., } $j\gg 1$). This leads to a
more accurate classical analog of the transition probability $\left( \text{%
\ref{a9}}\right) $, 
\begin{eqnarray}
W\left( {\bf p,p}^{\prime }\right) &=&\ \frac{\zeta \left( {\bf q-q}^{\prime
}\right) }{4\pi L^2m^2}\delta \left( \epsilon -\epsilon ^{\prime }\right)
\times  \label{ap2} \\
&&\ \left[ 2p_x^4\delta \left( p_x-p_x^{\prime }\right) +\frac{\Omega ^2}4%
\delta ^{\prime \prime }\left( p_x-p_x^{\prime }\right) \right] ,  \nonumber
\\
\Omega \left( {\bf p,p}^{\prime }\right) &=&\left( {\bf q}-{\bf q}^{\prime
}\right) \cdot \left( p_x{\bf q}+p_x^{\prime }{\bf q}^{\prime }\right) , 
\nonumber
\end{eqnarray}
than Eq.$\left( {\rm I}.18\right) $.

The corresponding change in the classical collision integral does not result
in any significant changes in the expressions for the classical transport
coefficients. The {\em only} improvement should be the substitution of the
functions

\[
\frac{d\,\sin \theta }{\alpha +4\tan ^4\theta } 
\]
in the integrands for {\em all} transport coefficients by 
\[
\frac{d\,\sin \theta }{\alpha +4\beta \tan ^4\theta +8\tan ^4\theta } 
\]
where, as in {\rm I}, $\alpha \left( u\right) =\left( 5/2\right)
\,_1F_1\left( 7/2,4,-u^2\right) /\,_1F_1\left( 3/2,2,-u^2\right) $, and $%
\beta \left( u\right) $ is similar, $\beta \left( u\right) =\left(
3/2\right) \,_1F_1\left( 5/2,3,-u^2\right) /\,_1F_1\left( 3/2,2,-u^2\right) $%
.

This change in the analytical expressions leads to more accurate results.
However, numerically, the change is not very significant. This small
numerical change is illustrated in Figs.6,7 for the functions $f_F\left(
x\right) $ and $f_B\left( x\right) $ which describe the transport
coefficients and the mean free path for Fermi and Boltzmann gases: 
\begin{eqnarray}
\sigma  &=&\frac{32}{\pi ^{3/2}}\frac{e^2L^2R^2N}{\hbar \ell ^2}xf_B\left(
x\right) ,\ x=\frac \hbar {\left( 4mT\right) ^{1/2}R},  \nonumber \\
\ f_B\left( x\right)  &=&x^4\int \frac{\exp \left[ -x^2z^2/\cos ^2\theta
\right] }{_1F_1\left( 3/2,2,-z^2\right) }\frac{dz}{\cos ^2\theta }\frac{%
d\theta }{\alpha +4\beta \tan ^4\theta +8\tan ^4\theta },  \label{ap3}
\end{eqnarray}
and 
\begin{eqnarray}
\sigma  &=&\frac{\sqrt{2}e^2L^2}{\pi ^3\hbar \ell ^2R}x^2f_F\left( x\right)
,\ x=\sqrt{2}\frac{p_FR}\hbar   \nonumber \\
f_F\left( x\right)  &=&\frac 1{x^3}\int \frac 1{_1F_1\left( 3/2,2,-x^2\cos
^2\theta \right) }\frac 1{\cos ^2\theta }\frac{d\sin \theta }{\alpha +4\beta
\tan ^4\theta +8\tan ^4\theta }.  \label{ap4}
\end{eqnarray}
Needless to say, all the prefactors remain exactly the same as in {\rm I}.
The variable $x$ describes the ratio $\lambda /R$ in Eq.$\left( \text{\ref
{ap3}}\right) $ and $R/\lambda $ in Eq.$\left( \text{\ref{ap4}}\right) $. 
\begin{figure}
\centerline{\input r_difig6.tex}
\caption [F6]{Function $f_F(x)$; solid line - Eq.(\ref{ap3}), dashed line - results I}
\label{fig.6}
\end{figure}
\begin{figure}
\centerline{\input r_difig7.tex}
\caption [F7]{Function $f_B(x)$; solid line - Eq.(\ref{ap4}), dashed line - results I}
\label{fig.7}
\end{figure}

\end{document}